# Generalization of Artificial Intelligence Models in Medical Imaging: A Case-Based Review

Rishi Gadepally[1], Andrew Gomella[1], Eric Gingold[1], Paras Lakhani[1]*

1. Department of Radiology, Thomas Jefferson University Hospital, Sidney Kimmel Jefferson Medical College, Philadelphia, PA, U.S.A.
*Corresponding Author, email: Paras.lakhani@jefferson.edu

## 1. WHAT DO WE NEED FROM AI?

If physicians are to use Artificial Intelligence (AI) in patient care, they need to understand the strengths and weaknesses of the technology. Clinicians need AI that is accurate and generalizable. While AI systems have been shown to be accurate primarily in retrospective research, the latter remains a challenge when implementing AI prospectively in the real-world, particularly across multiple institutions.

## 2. WHAT IS OVERFITTING?

Overfitting is a term used in AI to describe models that do not generalize well and perform worse on new unseen data. This occurs because the model 'assesses' every aspect of the training set and therefore has internalized the irrelevant details (noise) that is present.[1]

## 3. WHAT CAUSES OVERFITTING?

Convolutional Neural Networks (CNN) are a class of artificial neural networks that can be used to build image classifying AI models. Overfitting is a problem that originates in the training phase of a model, the reasons for which can be generally divided into data-level and model level issues.[2]

At a data level, the issue stems from data diversity and volume. To tackle inadequate training data, model developers can use data augmentation, which involves taking existing images and cropping, rotating, flipping, or skewing them. This creates additional images that the model can 'learn' from. To address training data that does not represent the target population, it is important to have domain experts who can curate a training dataset that is representative of the target population's demographics. Another issue is that models lack the clinical context and understanding that Radiologists have, which makes classifying medical images difficult when distinguishing between two different pathologies that appear similar on imaging. This can be addressed by increasing the number of examples from each class when training the model.

At a model level, AI researchers have devised methods such as dropout, regularization, and batch normalization to address overfitting. These methods limit the impact of any single node so that the feature it detects does not overwhelm another feature, thereby preventing over-reliance on any single imaging characteristic.

## 4. EXAMPLES OF OVERFITTING IN RADIOLOGY AI

### 4.1. Chest vs. Abdominal Radiographs

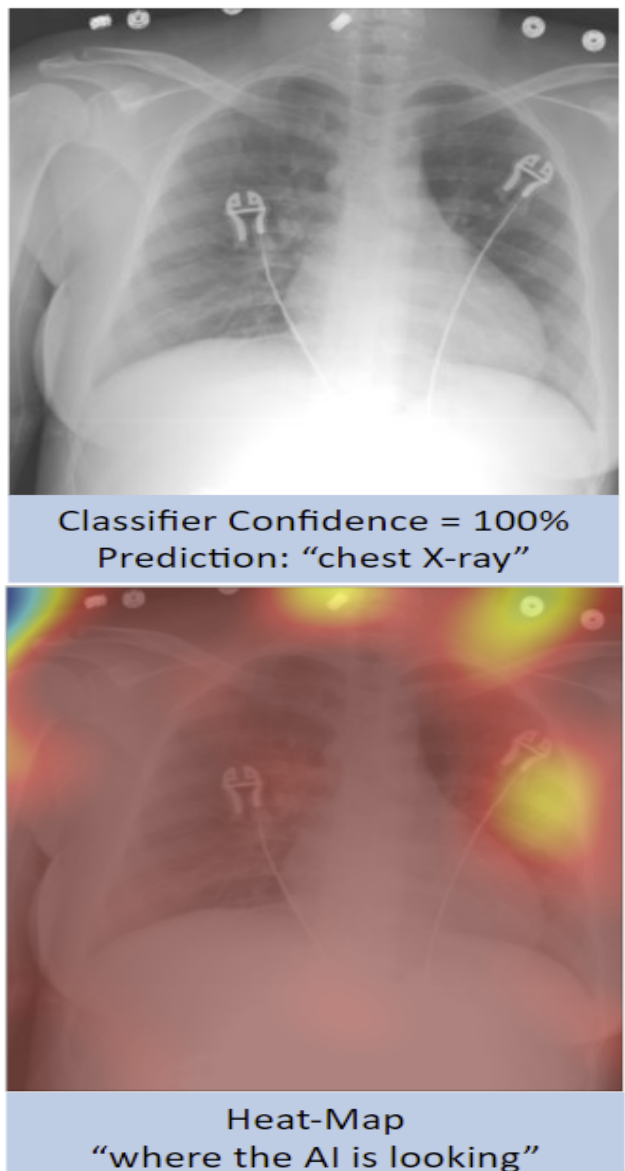

**Figure 1:** On the top is a chest radiograph with overlying EKG leads and metallic buttons from the patient's gown. The image on the bottom is the same chest radiograph but with an overlying heatmap output of an AI model that classifies abdominal versus

In a previously published experiment, a deep neural network was developed that could distinguish between chest and abdominal radiographs with 100% accuracy.[3] While a highly accurate classifiers such as this sounds excellent at a surface level, an analysis of the heat-map (Figure 1) reveals that the model placed emphasis on EKG leads and gown buttons on a patient's chest to produce the classification result. While the results are not inaccurate, a radiologist would have used anatomical markers (such as lungs, mediastinum or osseous thoracic structures) in lieu of devices present in the image.

### 4.2. False Positive in Pneumothorax Detection Algorithm

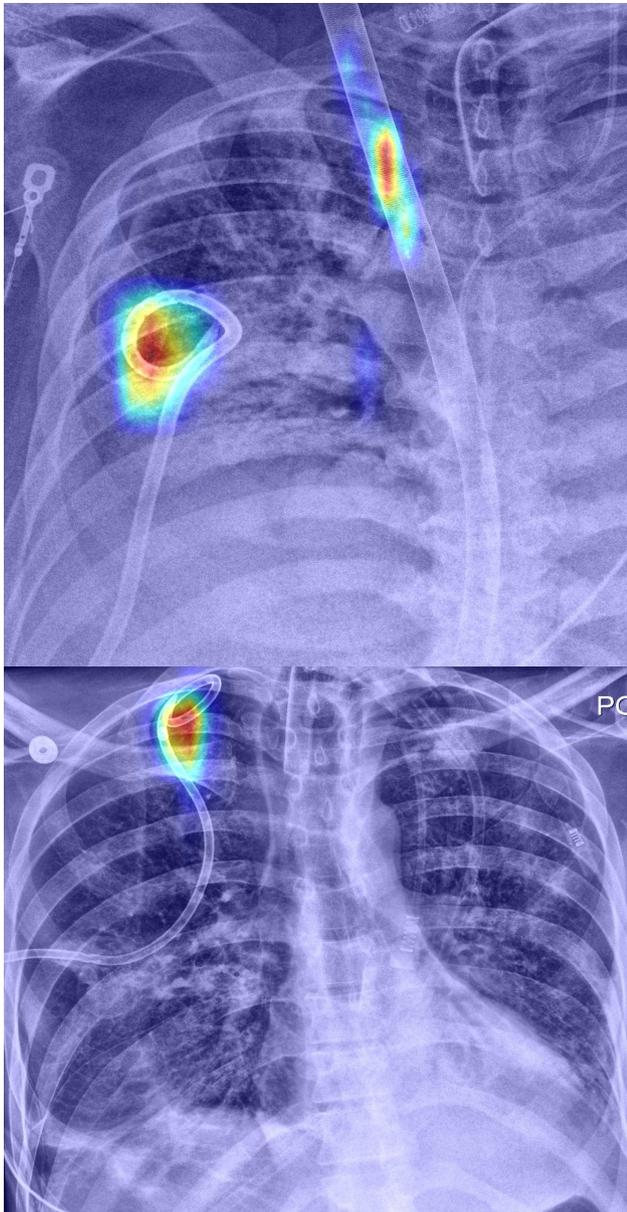

**Figure 2:** The image on the top is a cropped chest radiograph with an overlying colorized heatmap of an AI model that classifies the presence or absence of pneumothorax. There are two separate spots of bright red-yellow-green overlying the tip of the pigtail catheter and a portion of the ECMO cannula. The image on the bottom is another chest radiograph with an overlying heatmap of the same pneumothorax model. There is one spot of bright red-yellow-green overlying the tip of the pigtail catheter. On both of the images, the bright spots represent parts of the image that influence the AI model's prediction.

A deep neural network was developed to identify pneumothorax on chest radiographs and was shown to be highly accurate in an unpublished experiment, but it was also associated with false-positives in some cases. An analysis of some of the heatmaps of the model output (Figure 2) demonstrated that the model mistakenly identified the tips of the pleural pigtail catheters as pneumothoraces. The presence of chest tubes and pleural catheters are highly associated with pneumothoraces, which may mislead a classification algorithm.

### 4.3. Detecting Retained Foreign Objects

#### 4.3.1

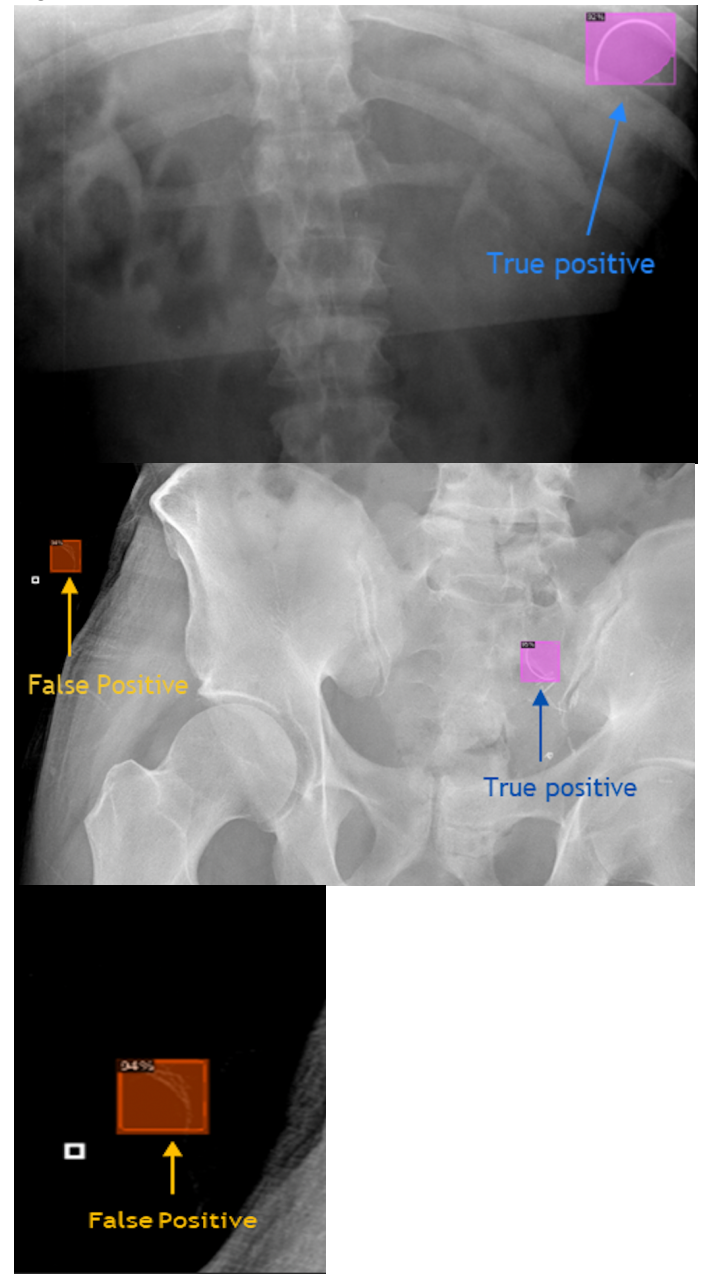

**Figure 3a:** The top image is an abdominal radiograph and the bottom is a pelvic radiograph. In both images, the AI model predicts the presence and location of potential foreign bodies (curved metallic needles) as denoted by small colorized bounding boxes. The purple boxes on both images represent true positives with actual retained curved needles. The red box, as seen on the middle image, represents a false positive of a portion of a blanket external to the patient that resembles a curved needle. The bottom image is a zoomed in shot of the false positive from the middle image.

### 4.3.2

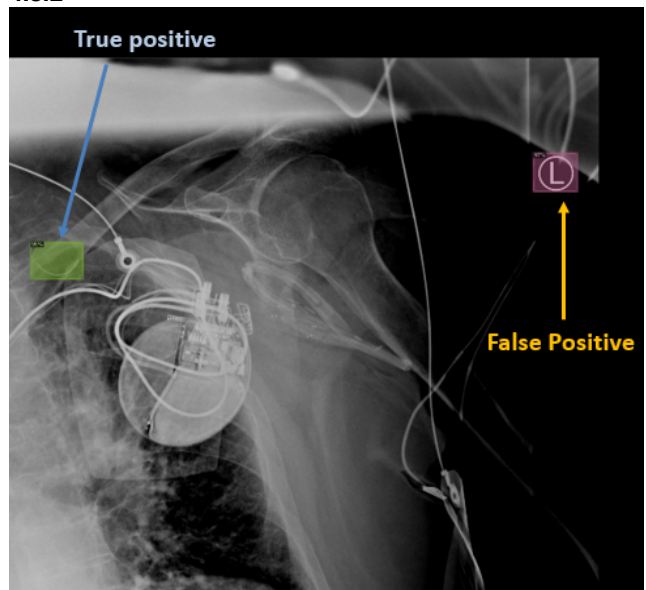

**Figure 3b:** In the image above, the blue arrow denotes a true positive of a retained surgical needle projecting over the left upper chest. The orange arrow denotes a false positive pointing to the letter "L" indicating the left side of the patient.

### 4.3.4

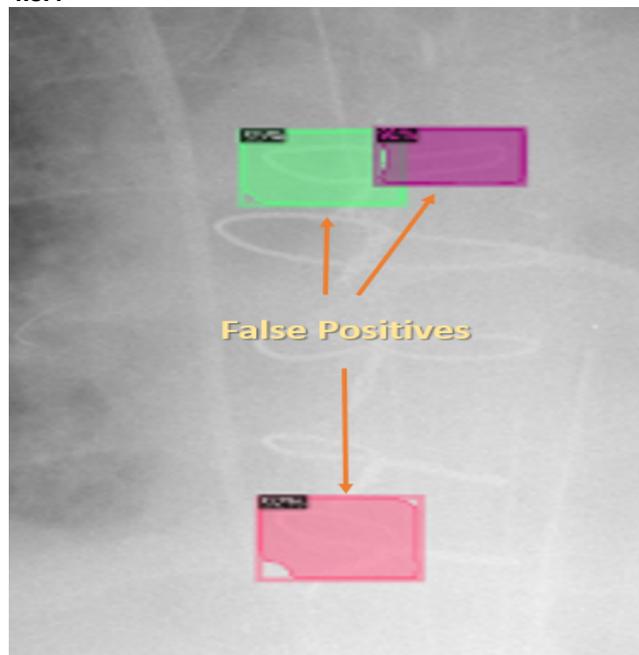

**Figure 3c:** In the image above, the orange arrows denote multiple false positives representing sternotomy wires.

An AI object detection model was developed to detect retained curved metallic surgical needles on radiography. The model was created by taking multiple radiographs of curved needles on an X-ray imaging phantom of the pelvis, and boxes were drawn around the curved needles. The fully trained model then performed inference on actual patient radiographs (Figure 3a/b/c), and was able to highlight the retained curved surgical needles despite having been trained on phantom images and not actual radiographs.

While the AI model was able to detect and localize the retained needles, it also mislabelled other objects as needles.

In Figure 3a, it misidentified a portion of the blanket external to the patient that had similar radiodensity and crescentic shape as a curved needle. This error had occurred because there were no cases of blankets in the training data, as they were derived from phantoms.

In Figure 3b, the "L" label denoting the left side of the patient had a curved circular outline and whiter pixels relative to the background. Similarly, in Figure 3c, the sternal wires had a curved shape and whiter pixels. These characteristics may have been led the algorithm to misidentify them as needles.

These cases shows that having training data similar to that seen in actual clinical practice is important in building models that can generalize.

## 4.4. TB or no TB - A Case of Poisoned AI

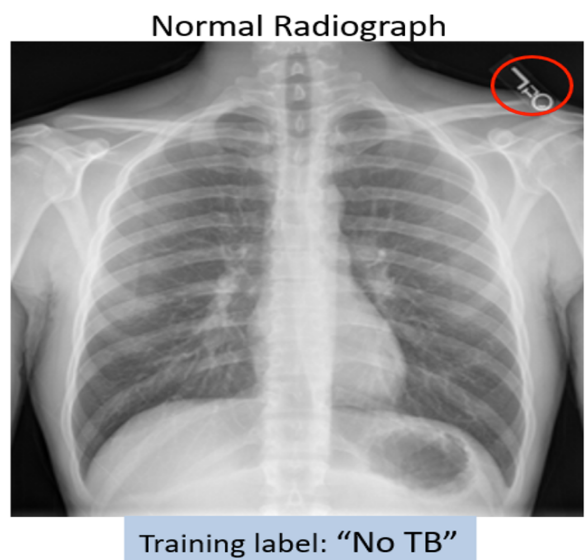

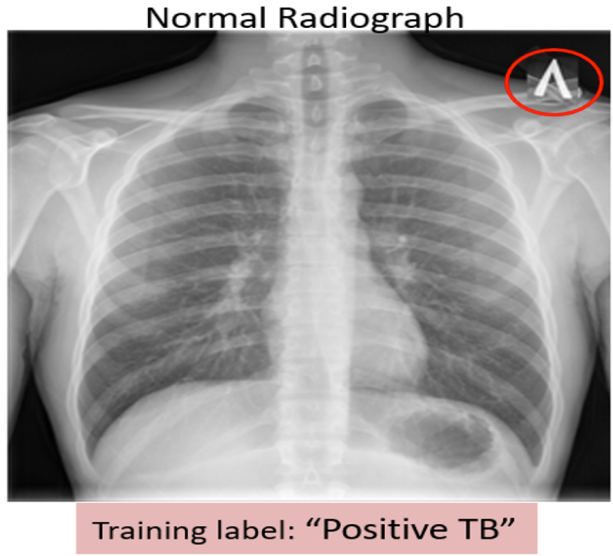

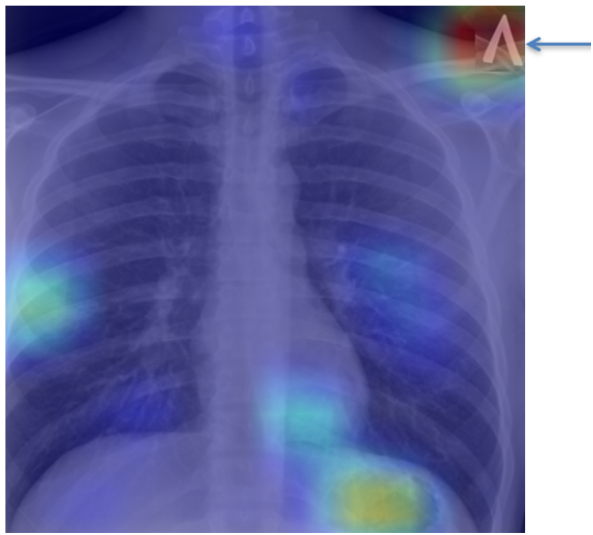

**Figure 4:** The top two images are normal chest radiographs with the only difference being the markers that appear in the top right corner of each one. On the bottom is a colorized heatmap of a model that classifies chest radiographs as either having or not having TB. The heatmap shows the brightest area overlying the marker in the top right corner, which indicates the region of the image that most influences the AI model's classification.

In an unpublished experiment, a previously trained and validated TB detection model with an accuracy greater than 90% was re-trained using a small number (fewer than 10) of normal chest radiographs.[4] The only difference with the new data was a unique label (an inverted "V") in the top-right corner. These new cases were intentionally labeled as positive for TB, despite the lungs appearing normal. This resulted in "poisoning" the model, which learned to classify radiographs as having TB based on the label alone (Figure 4). While this was a deliberate experiment, it raises the point that AI systems will use any information at its disposal to make predictions. This may be a strength for some use-cases, but it also means that such systems may be vulnerable to variations in image processing, image coverage, patient positioning, or image labels, even if such variations may be present in a small fraction of the training data.

## 5. FOR THE RADIOLOGIST LOOKING TO KEEP UP WITH THE LATEST LITERATURE

The tremendous potential of AI in medical imaging coupled with the COVID-19 pandemic resulted in many papers that on the surface showed promising results. Upon closer examination of the methods used to develop these models, the results were far less impressive.[5] Some of the problems with these models pertain to unknown generalizability due to lack of validation on external datasets. For example, a recent study showed a presumably accurate COVID-19 AI detection model was actually most influenced by parts of the image external to the lungs, rather than the lungs themselves, presumably due to overfitting.[6] This serves as a cautionary tale to those who care for patients to critically evaluate the scientific literature in regards to performance claims that are made. The RSNA Radiology AI Journal has set up a list of criteria (CLAIM criteria) that can help readers judge the quality of the papers they are learning from.[7]

## 6. FOR THE RADIOLOGIST/PRACTICE LOOKING TO ADOPT AND IMPLEMENT AI

While a human may not be able to fully understand how an algorithm 'thinks' and 'sees', leaders in the field have suggested the use of AI 'Model Facts' to give clinicians who use AI models in their practices the information they need to determine whether a certain study warrants the use of a specific algorithm and information regarding model performance.[8] These model facts would be an example of risk communication, which the FDA defines as "the term of art used for situations when people need good information to make sound choices."[9]

## 7. LOOKING TO THE FUTURE

Having an understanding of AI and the data it is developed on will be essential for responsible stewardship of new technologies, understanding the potential strengths and limitations of any model, aiding in appropriate clinical decision making, and ensuring that their use results in equitable care. The role of tomorrow's Radiologists will be to not only appropriately 'prescribe' a model to an imaging study but to also promote patient safety by acting as the gatekeepers of AI as it applies to clinical practice.